\documentclass[9]{acm_proc_article-sp}
\pdfoutput=1
\usepackage{url}
\usepackage{float}
\usepackage{setspace}
\usepackage{verbatim}
\usepackage{tabularx,ragged2e} 
\usepackage{appendix}

\begin{document}

 \title{Semantic Identification of Web Browsing Sessions}

\numberofauthors{1} 
\author{
\alignauthor
Neel Guha \\
\affaddr{Stanford University}\\
\affaddr{CA, USA}\\
\email{nguha@stanford.edu}
}

\maketitle

\begin{abstract}
We introduce a \textit{semantic identification attack}, in which an adversary uses semantic signals about the pages visited in one browsing session to identify other browsing sessions launched by the same user. Current user fingerprinting methods fail when a single machine is used by multiple users (e.g., in cybercafes or spaces with public computers) as these methods fingerprint devices, not individuals. We demonstrate how an adversary can employ a SIA to successfully fingerprint users on public or shared machines and identify them across browsing sessions. We additionally describe and evaluate possible countermeasures to prevent identification. 
\end{abstract}

\section{Introduction}
Online privacy is becoming an increasingly important issue for users, policy makers, and academic researchers. In recent years there has been significant work highlighting the fragility of user privacy by exposing threats ranging from the de-anonymization of public data sets to third party tracking in online browsing.  A particular area of focus is user fingerprinting in web browsing. Prior work has demonstrated that by analyzing device and network information (screen resolution,  extensions installed,  IP address, etc), an adversary can uniquely identify an individual and track their browsing across multiple sessions. 

However,  because these methods rely on information about the device, they effectively fingerprint devices and not users. If two individuals use the same device (leaving the configuration and settings unchanged), current fingerprinting techniques would fail to distinguish between the two individuals. An adversary may incorrectly assume that there is one user whose activity is the composite of the two individual's actions. 

Though web browsing devices are considered ubiquitous today, it is still common for a single device to be shared by multiple individuals. Many homes have a single computer that is shared by a family.  Spaces such as libraries or lounges often have publicly available devices. In developing countries, internet cafes are common for individuals who can't afford personal computers.

The primary contribution of our work is to show that users can still be identified in these cases. Our intuition is that a user's behavior can provide the basis for that user's fingerprint. We introduce methods capable of fingerprinting users by their behavior in web browsing.  Specifically, we present a \textit{Semantic Identification Attack}, in which an adversary uses the semantic signals from a user's browsing session (i.e. the content of the pages they visited) to identify other sessions launched by the same user. A user is unlikely to visit the same urls in every session. However, the urls in a given user's browsing sessions are unlikely to be drawn completely at random from all the urls on the web. If a user's web browsing is characterized by a set of tasks (checking certain news topics, reading specific blogs, etc), their browsing sessions will consistently contain visits to pages/websites relevant to these tasks. The distribution of these visits can be viewed as a user specific session fingerprint. Though a user's browsing sessions are likely to deviate and contain other behavior as well, the presence of these pages fingerprints the user to the session. 

User fingerprinting can be both beneficial and harmful. Law enforcement could use fingerprinting to track and identify individuals engaging in harmful or malicious activities. Alternatively, authoritarian governments could use fingerprinting to track dissenting individuals and suppress citizens. 

The outline of the paper is as follows. We first review past work in user fingerprinting. We then provide an example and describe the optimum threat model for a semantic identification attack. We describe various methods for launching the attack and validate each of them on several different browsing datasets. Finally, we discuss and evaluate several countermeasures a user could employ to protect their privacy.

\section{Related Work}

There has been significant work on fingerprinting users on the basis of their browser\cite{Yen2009}\cite{mowery2011fingerprinting}. \cite{Eckersley2010}  and \cite{host-fingerprinting-and-tracking-on-the-webprivacy-and-security-implications} describe how users can be uniquely identified by analyzing system fonts, screen resolution, information stored in HTTP headers (ie, HTTP User Agent strings) and coarse IP prefix information (users can test their fingerprints at \cite{pano} or \cite{amiunique}). Additionally, \cite{MS12} showed that it is possible to fingerprint users by leveraging the subtle differences in the ways different browsers render the same text. These differences provide consistent, unique fingerprints that  can be used to track users.

However, there has been little work on identifying user web browsing without using browser metadata or network information as features. Our paper shows however, that even if a user were to take these precautions, an adversary could still identify them on the basis of the content of their browsing. Browsing history is a strong predictor for learning user attributes, suggesting that semantic signals from browsing sessions could convey significant discriminating information \cite{goel2012does}.  Jones et al \cite{Jones:2007:IKY:1321440.1321573} show that it is possible to infer personal information (gender, age, location, etc) from query logs and link a particular user to a query stream when provided with some background information on that user.  \cite{Kumar:2007:AQL:1242572.1242657} demonstrates that even a countermeasure like token-based hashing of queries fails to prevent an adversary from identifying users. \cite{5708146} proposes  the ZEALOUS algorithm, which is capable of achieving strong privacy guarantees on query logs.  \cite{Novak:2004:AW:988672.988678} focuses on determining when a single individual is masquerading behind multiple aliases on a web forum using the  content present in forum posts. It is similar to our work in that it aims to use semantic information to uniquely identify individuals. However, \cite{Novak:2004:AW:988672.988678} is applied to a forum (as opposed to web browsing) and leverages significantly richer data. 

Another closely related field of study is differential privacy\cite{narayanan2010myths}. Though differential privacy is primarily concerned with the privacy guarantees of statistical databases and their individual entries,  some of the core concepts and ideas are applicable in our case. \cite{Dwork2008} describes some key findings. \cite{agrawal2000privacy} also present techniques for preserving privacy in data mining applications. The 2006 AOL scandal demonstrated that there are techniques available to identify users from anonymized web search data \cite{aol_scandal}. Narayanan and Shmatikov presented a further class of de-anonymization techniques by using the IMDB dataset to identify individual users in the Netflix Prize Dataset \cite{Narayanan:2008:RDL:1397759.1398064}.

\section{Threat Model}

We now describe the threat model for a semantic identification attack. Specifically, we present the optimal case for such an attack, discuss the damage an adversary is capable of, and describe how traditional attacks (discussed previously) may fail.  

We focus on a single computer used for web browsing at an internet cafe. Different individuals would use this machine to launch browsing sessions, during which they visit various web pages of interest to them. Individuals repeatedly use this machine, so that over a period of time (perhaps several weeks), a majority of individuals would have launched multiple browsing sessions on this machine. An adversary launching a semantic identification attack on this computer would seek to identify which browsing sessions belonged to the same user.  Ideally, the adversary would be able to group all browsing sessions by their users. We say that an adversary is able to "successfully fingerprint" a user if they are able to identify the exact set of sessions belonging to that user. 

This threat model makes the following assumptions on the adversary's capabilities:

\begin{enumerate}
\item An adversary could be any attacker that has access to at least some fraction of the user's session logs. At the very least, the adversary needs to know some information about the content of at least some of the pages  visited by the user in a session. The adversary also needs to be able to distinguish between different sessions (i.e., know when one session ends and another begins). If the adversary knows the URLs visited by a user, we assume the adversary can crawl them later. 
\item The adversary cannot physically "see" the user. The problem of identification could be solved trivially if the adversary were to sit in the cafe and observe every user who used the device. We can safely assume that the adversary is operating remotely and has no way of physically seeing which users are browsing on the device at any given time. 
\end{enumerate}

Given the flexibility of the requirements, the scope of potential adversaries is quite large. An analytics or advertising network could launch this attack using third party cookies. These cookies would allow the adversary to track the various the web pages that a user visits within a single session. Equivalently, a website like the New York Times - where multiple users may use the same account to access content - may use this attack to determine which browsing sessions belong to the same user (despite all sessions being in the same "account name").

Despite the perceived ubiquitousness of personal web browsing devices (phones, laptops, etc), the threat model described above is widely applicable. Internet cafes are incredibly popular in developing countries, where individuals cannot afford to purchase a personal computer but still need to perform various web tasks. For example, a 2013 South African study revealed that 9.6\% of South Africans who go online do so through internet cafes \cite{rsa}.  As of July 2012, 25.4\% of internet users in China frequently visited an internet cafe (138.3 million users) \cite{china_ic}.  

The threat model described above could be generalized for any shared device. This could include homes/small businesses that own one computer, or any publicly available computer (like those found in lounges, libraries, airports, etc).  

Because most fingerprinting techniques rely on device specific metadata (browser configurations, network information) they instead fingerprint devices and assume that every device has a single user. By not using device metadata and instead relying on the content of web pages visited by users, a  semantic identification attack does not require the one-device-one-user assumption. In this paper, we show how such an SIA can successfully fingerprint  users within the threat model presented above and evaluate various approaches on  browsing data.

\section{Content Fingerprint}

As mentioned above, a semantic identification attack attempts to identify and distinguish users based on their behavior (the pages and websites they visit). In order to do so, the attack aims to construct a fingerprint for each user from their behavior in a single session. We  refer to this fingerprint as a user's \textit{content fingerprint}.  We now describe the motivation for this behavior based fingerprint. 

\subsection{Intuition}

The goal of a SIA is to distinguish between multiple individuals browsing the web on the same machine. Web browsing is a highly personal activity and the pages a person visits are a function of their interests, tasks, and habits. Observe that every page can be described as a set of \textit{semantic signals}, or features which broadly describe the content of the page (ie the words on the page, entities mentioned, etc). By taking all pages visited by a user, we can observe a distribution over all semantic signals that describes the user's interests/tasks/habits. We refer to this distribution as a user's \textit{profile}. Similarly, we can observe the distribution of semantic signals over a single session of browsing. This distribution is the user's \textit{content fingerprint}. A semantic identification attack makes two primary assumptions: 

\begin{enumerate}
\item \textbf{Uniqueness}: We assume that an individual's profile is unique. Since a SIA attempts to distinguish the individuals using a single machine to browse the web, the set of users being compared is relatively small. This assumption of local uniqueness is thus relatively safe. 

\item \textbf{Consistency}: We assume that the content fingerprints for two distinct sessions from the same user are approximately equal. Since a user's profile is a distribution across all pages visited, the content fingerprint is effectively a sample from this distribution (corresponding to a single session). We can thus safely assume that a user's sessions are representative of their profile. 
\end{enumerate}

Exceptions to these assumptions are discussed in further depth later in this paper. 
Because a user's profile is unique and their browsing consistent across sessions, it should be possible to fingerprint users through their content fingerprints. More precisely, given two sessions, an adversary should be able to compare the content fingerprints of each in order to determine if they were launched by the same user. If the content fingerprints are sufficiently similar, then an adversary can safely decide that they share the same user. If they are different, then an adversary can decide otherwise. Thus, the goal of the adversary is to identify semantic signals from each page and extract content fingerprints from each session such that the two assumptions above are satisfied. If the content fingerprints extracted from each session aren't granular enough, then content fingerprints will not be unique. However, if content fingerprints are too granular, then it will be impossible to resolve that two sessions have the same user, as their content fingerprints will look too different.

In the following sections we formalize this intuition and explain this process in further detail using an example. 

\subsection{Formalization}

Every page has an associated set of features (semantic signals). A user profile $b$ is the distribution over these features for all pages visited by a user $u$. Given a session consisting of a set of $s$ pages visited by $u$, we expect the distribution over these features - the content fingerprint, given by $f$ - for the session to resemble $b$ as $s$ grows. 

Thus, given a session $s_1$ from a user $u$ with a browsing profile $b$, we can say that a session $s_2$ belongs to $u$ if 
\begin{enumerate}
\item $b$ is unique amongst all users, and 
\item Both $s_1$ and $s_2$ have sufficiently similar distribution over features. Ie, $f_1$ (corresponding to $s_1$) is sufficiently similar to $f_2$ (corresponding to $s_2$)  
\end{enumerate}

If either of these conditions is false, then we cannot definitively say that $s_1$ and $s_2$ belong to the same user. 

When extending this to multiple users, we do as follows. Take a set of sessions $S = \{s_1,...,s_m\}$ where each $s \in S$ was launched by a user from the set $U = \{u_1,...,u_n\}$ but the mapping from $s_i$ to $u_j$ is unknown (ie, we do not know which sessions were launched by which users).  An adversary launching a SIA would attempt to partition $S$ into $n$ subsets  such that all the sessions in subset $i$ were launched by $u_i$. For each session, the adversary would attempt to construct a content fingerprint (giving $f_1,...f_m$). Since content fingerprints are unique to users, the adversary would compare all $f_i$. If two sessions had similar  $f_i$, then the adversary would know that both originated from the same user. 

\subsection{Semantic Signals}

The content fingerprint of a user is a representation of that user's interests, habits, and tasks. In order to capture these, the adversary needs some signals regarding the topics/content of the user's browsing. We refer to these as the semantic signals from the session.  Semantic signals capture the content of each page at some level of granularity. They allow the adversary to extract the user based content fingerprint. Some examples of semantic signals include:

\begin{itemize}
\item The words on the page visited
\item Categorical classifications for the page visited (e.g., "news", "social media", "cooking")
\item The title of the page
\item Links on a page
\item URL of a page
\item Page tags (derived from social bookmarking sites, knowledge bases, etc)
\end{itemize}

The success of a semantic identification attack is highly dependent on the granularity of the semantic signals. Rich signals - such as a listing of every word on every page visit will allow the adversary to extract a more powerful content fingerprint. Weak signals - such as a simple classification of each web page (ie "news"), is rather poor and will lead to fewer unique content fingerprints. 

The semantic signals available to an adversary depend on the adversary's capabilities and the position from which they launch the attack. In this work, we demonstrate that even a weak adversary with poor signals can still uniquely identify some users. 

\subsection{Practical Considerations }
We now discuss several practical considerations of SIA attack strategies.  

\subsubsection{Fingerprint Uniqueness  \label{limitations}}
An SIA assumes that an adversary will be able to extract a content fingerprint from a session that is unique enough to identify the user. This is not always possible. If a user's behavior in a session doesn't sufficiently reflect their user profile, then it will be difficult for the adversary to extract a representative content fingerprint. This could occur when 

\begin{enumerate}
\item The session is not long enough, and contains relatively few page visits. 
\item The session represents a one time task like planning a vacation or reacting to an emergency
\item The user's behavior is not unique.
 \end{enumerate}
The first case is analogous to a user's session consisting of a relatively small sample from their profile. Unless the sampled sessions contains semantic signals that are highly unique to the user, it'll be much harder to determine the owner of the session. 

In the second case, the user's behavior deviates significantly from their profile. Most users do not routinely plan vacations or deal with emergencies - these are one
time tasks. As a result, semantic signals from these tasks are unlikely to be represented in the user's profile. Thus, resolving these sessions to the user will be a difficult. 

In the third case, the user's profile is not sufficiently unique. For example, it's probable that there are many individuals who use the web only for "general" or "common" tasks. For example, their web browsing may be restricted to checking the weather and reading the local news. These tasks aren't very personal or unique. It will be nearly impossible for an adversary to extract discriminating and consistent content fingerprints for each of these individuals.  

It is important to recognize the limits of this attack. While it is probable that any  two randomly selected individuals will differ significantly in their browsing behavior, identifying users grows in difficulty as the population of users increases. Intuitively, an SIA attempts to identify distinguishing characteristics between users. Since larger populations of individuals are more likely to converge on some "median" behavior, identifying these characteristics becomes relatively harder with more users. Though an SIA can be an effective attack (as we demonstrate), it is successful only on a proportion of the users. There will be certain users - namely those who lack any unique behaviors - for whom the attack will likely fail.

\subsubsection{Fingerprint Similarity}
Even if two sessions were launched by the same user, it's unlikely that the extracted content fingerprints will be exactly the same. Thus, determining whether two sessions share the same user should be based on the fingerprint similarity of the two sessions. If the fingerprints are similar, then we have more confidence that they originated from the same user. 

\subsection{Example}

We now describe an example to illustrate the principles described above.  

Imagine a farmer living in rural India. Several times a week he visits an internet cafe and spends an hour browsing the web. As a farmer, he is interested in the local weather patterns and the markets for his produce (focusing specifically on certain types of goods). He may also visit various religious blogs to find out about upcoming festivals. All of these topics act as markers for this user. His content fingerprint  would likely indicate a skew towards pages covering these topics. He is characterized by these interests - the locality of his news, his professional interests, and his religious interest. Individually, none of these interests are unique and are likely shared by others frequenting this cafe. When combined however, the set of users who lie at this intersection is much smaller. 


\subsection{Fingerprint Granularity}
Given a browsing session, we can either extract a single fingerprint for the entire session (a cross site fingerprint) or multiple fingerprints, one for each website in that session.  Most web browsing sessions consist of page visits to multiple different websites. If we extract a cross site fingerprint, then we extract a single fingerprint that describes the user's browsing behavior over all of these websites. If we extract site fingerprints, then for each website in the session we extract a single fingerprint describing the user's behavior on that website in that session. Because these fingerprints are being extracted on a per site basis, we extract multiple fingerprints per session (one for each of the websites visited). As each of these fingerprints reflect more localized behavior (on a single website), they are less likely to be unique across users. However, we can use the aggregation of fingerprints across websites to identify users.

Though individual website based fingerprints are less insightful, they are more resistant to user defenses. A cross site fingerprint for example, may be hard to collect if the user blocks third party cookies (preventing cross site  data from being collected by an adversary). However, the website a user visits will always know the pages clicked on by the user. If this website can extract a granular site specific fingerprint for the user, they can use that to identify the user when they revisit the website. Thus, even if a user uses incognito mode to make these visits, the website may be still be able to track them.

In this paper, we analyze both types of fingerprints. Our Delicious dataset is an approximation of user browsing behavior from multiple sites, thereby allowing us to create a cross-site fingerprint. In contrast, our MSNBC browsing dataset is restricted to web browsing data on msnbc.com, forcing us to develop a website specific fingerprint. As we later demonstrate, we are able to launch a successful attack using both types of fingerprints. 

\section{Empirical Validation}
In order to validate our attack strategies, we evaluated them on real browsing data. Unfortunately, the lack of datasets of real browsing sessions  severely constrains our ability to test different attack strategies. Before we present our strategies, we briefly describe the two datasets we use in this paper. 

\subsection{MSNBC Browsing Dataset}

We used the MSNBC Anonymous Browsing Data set (\cite{msnbc_data}). The data contains a list of pages visited on msnbc.com for 990,000 users over a 24 hour period (on September 28, 1999). For each user, we are given a list of the pages they visited. However, pages are only represented at a "category" level (eg, "homepage", "sports", "news", etc). 

\subsection{Delicious Bookmark Data}

We used a Delicious bookmark dataset to approximate user browsing \cite{delicious}. Delicious is a social bookmarking service, which allows users to "bookmark" web pages with descriptive tags. The initial dataset contained 69226 pages from 1867 users. We were able to successfully scrape 40171 pages from 1850 users. Restricting our attention to these pages, we were able to to acquire a range of semantic signals (which we discuss in depth later). 

\section{Attack Strategies}

We launch/evaluate our attacks in the following manner. We begin by sampling $n$ users, where each user has visited a minimum of $k$ pages. We randomly permute the order of these $k$ pages, and divide them into two sets. Each set corresponds to a browsing session. Thus, for $n$ users, we will create a set of $2n$ browsing sessions. Given this set of sessions, the goal of the attack is to identify which pairs of sessions belong to the same user. We evaluate the success of the attack using the metrics discussed above.   

We now discuss the attack strategies we employed. We can think of this as compromising of two primary tasks: fingerprint formulation and session similarity scoring. In fingerprint formulation, we seek to determine some way of representing each session's content fingerprint. In session similarity scoring, we seek to identify ways of determining if two content fingerprints are sufficiently similar to have belonged to the same user.

\subsection{Fingerprint Representation}

We now describe the features used in representing the fingerprints for both sets of data. 

\subsubsection{MSNBC}
Unfortunately, because of the restrictive nature of the MSNBC dataset, we do not have much scope to explore fingerprint representation. In the dataset, each page is represented as a page category (a page can be "news", "homepage", "local", "tech", etc). The granularity of our data is restricted to the page category, for a given page that a user visited, we only know what category the page falls under. We have no other information about that particular visit. Below is a sample representation of the data: \[
\begin{bmatrix}
     \{1,3,5,1,3,4,4\} \\
    \{1,3,7\} \\
    \{5,7,9,14,15\} 
\end{bmatrix} \]
Each line in the matrix above corresponds to the browsing activity for a single user. For example, in this case the data tells us that user corresponding to the second line visited a page of category type 1, followed by category type 3, and finally category type 7. This user visited no other pages on msnbc.com during this time period. 

These categories are the "semantic signals" for each of the user's page visits. Despite the coarse grained information(very different pages could be labelled with the same category), we show it is still possible to construct unique fingerprints for a subset of users and to correctly identify sessions originating from them.  We represent each session as a vector of the proportions of each of the 17 categories of pages visited in each session. Thus if a session contained the visits to the following pages: $$[1,2,2,2,2,3,4,3,3,3]$$
The corresponding vector representation would be $$[0.1, 0.4, 0.4, 0.1, 0,0,0,0,0,0,0,0,0,0,0,0,0]$$
Since visits to pages of category "1" represent 10\% of the session's page visits, pages of category "2" represent 40\% of the session's page visits, and etc. Each session can be seen as a unit vector in a 17 dimensional space, where each dimension corresponds to a category. 

This representation is general and intuitive, but does fail to capture some potentially interesting information such as the order in which the user looks at pages in different categories. 

\subsubsection{Delicious}
For the delicious data, we formulated several different ways of representing session fingerprints. 

\begin{enumerate}
\item \textbf{TFIDF}: We treat each session as a single "document" by collating all words across all pages in that session. We represent each session as the vector of the TFIDF scores of its words (using all sessions as the document corpus). We ignore all words that appear in more than 95\% of documents and we remove stop words. We use a single layer autoencoder to reduce the dimensionality of the resulting vector.  
\item \textbf{Domain}: We represent each page by the domain name. Thus, the content fingerprint of a browsing session is represented as a single vector, where each dimension corresponds to a different domain name and the value at an index is the number of pages in the browsing session that have the respective domain name. 
\item \textbf{Delicious Tag}: We represent each page by the delicious tags associated with it. Similar to above, the content fingerprint is represented as a single vector where each dimension corresponds to a tag, and the value of the dimension is the number of pages which have the tag.

\end{enumerate}
We now discuss several different strategies for leveraging these fingerprint representations to determine when two sessions belong to the same user.

\subsection{Weighted Closeness}
If we can map sessions into a vector space based on their content fingerprints, two sessions belonging to the same user should be close to each other and far from sessions belonging to other users.  In measuring these relative distances, we can determine if two sessions belong to the same user. In this approach, we compute a score for every pair of sessions. If the score for two sessions exceeds a threshold, then we predict that two sessions originate from the same user. 

For two sessions represented as the vectors $s_i$ and $s_j$, we calculate $Score(s_i,s_j)$ in the following manner. 
$$ Score(s_i, s_j) = \dfrac{Cosine(s_i,s_j)}{\sum_{k=0}\dfrac{1}{Sim(s_i,s_k)} + {\sum_{k=0}\dfrac{1}{Sim(s_j,s_k)}}} $$where $Cosine$ is cosine similarity. This score is the combination of a similarity score  and a dissimilarity score. We discuss both in more detail below. 

\subsubsection{Similarity Score}
For calculating the similarity between two sessions $s_i$ and $s_j$, we use the cosine similarity metric, a common measure in information retrieval. Given two vectors in an n-dimensional space, cosine similarity measures their distance as a function of the angle between them. 
$$ Cosine (s_i, s_j) = \dfrac{s_i \cdot s_j}{\vert \vert s_i \vert \vert \vert \vert s_j \vert \vert} $$
We expect that two sessions belong to the same user will have approximately similar fingerprints, and thus a high cosine similarity score. 

\subsubsection{Dissimilarity Score}
Simply measuring whether two sessions are similar is not sufficient enough to capture whether or not they may originate from the same user. Our underlying goal is to be able to capture and compare the content fingerprints of two sessions. Thus, we need to weight the similarity of two sessions by how dissimilar those sessions are from all other sessions. If we have two sessions $s_i$ and $s_j$  such that $s_i$ is similar to $s_j$ but both $s_i$ and $s_j$ are similar to the bulk of the sessions in our data set, we are less confidant that $s_i$ and $s_j$ originate from the same user. it is probable that $s_i$ and $s_j$  (and the sessions they are similar to) belong to a mass of  users whose behavior is too shallow or generic to discern. Conversely, if $s_i$ and $s_j$ were similar to each other but different from other sessions, we would be significantly more confidant that both sessions originated from the same user. 

\subsection{Supervised Learning}

Alternatively, as opposed to devising our own similarity score, we can rely on machine learning to learn some scoring function. We rely on a multi-layer perceptron classifier from the scikit-learn python machine learning library\cite{scikit}.  For two sessions $s_i$ and $s_j$  with the feature vectors $c_i$ and $c_j$, our classifier takes as input the $\vert c_i - c_j \vert$  (the absolute value of the difference between the features).

\subsection{Metrics}
We now define the metrics used to evaluate the success of an SIA strategy. We consider our attack in the following framework. Assume we have a set of browsing sessions $S = \{s_1,s_2,..,s_n\}$ where each $s_i$ was launched by a single user and there is at least one other $s_j \in S$ such that $s_j \neq s_i$ and both $s_j$ and $s_i$ were launched by the same user. For every pair of sessions in $S$ the goal for the adversary is to identify whether they share the same user. 
We can measure an adversary's success along two dimensions. 

\subsubsection{Precision}
Precision is defined as the proportion of session pairings identified by the adversary that are correct. Let $S_p$ be the set of pairs of sessions such that for each pair the adversary has predicted both sessions originate from the same user. Let $S_t$ be the set containing all pairs of sessions in $S$ that originate from the same user. We can then formally define:
$$\text{Accuracy} =  \dfrac{S_t \cap S_p}{S_p}$$
\subsubsection{Recall}
We define recall as the proportion of same origin session pairs successfully identified by the adversary. More formally: 

$$\text{Recall} = \dfrac{S_t \cap S_p}{S_t} $$ 

\subsubsection{Reach}
We define reach as the number of unique users for whom the advertiser has successfully paired at least two sessions. Reach captures the "user footprint" of the semantic identification attack and allows us to quantify the number of users (not sessions) that an adversary has successfully compromised. 

\subsubsection{Success}

It is important to note that in this framework, there are many possible definitions for a "successful" strategy. In the ideal case, an adversary would prefer an attack strategy which grantees high accuracy and high reach - identifying sessions from a lot of users with  high probability of correctness. However depending on an adversary's goals, different dimensions may be prioritized. An advertiser for example, seeking to broadly identify users in order to show related advertisements, may care less about accuracy and more about reach. Since ads are generally viewed as terrible, they may care more about maximizing the number of users they reach (as opposed to the proportion of time they're correct). 

Conversely, a different adversary may care more about maximizing their accuracy (as opposed to their reach). If they're trying to closely track a set of users, they may care more about the strength of their predictions and less about accuracy.

In the empirical evaluations conducted in this paper, we choose to report to the precision and recall scores that correspond to the maximum F1 score achieved for an attack strategy.  Since different adversaries have different priorities, this metric conveys how well the "average" adversary would perform if employing the respective attack strategy.

\section{Defense Mechanisms}

\cite{rivest1998chaffing} presents a cryptographic scheme in which fake packets of information are interspersed with a sender's message in order to ensure security without requiring encryption. In this scheme, the receiver uses an agreed upon message authentication code to discard the fake packets from the real packets of information. An adversary intercepting these packets however, will be unable to distinguish the genuine data packets from the garbage ones, thus guaranteeing  security. 

We present a similar scheme to enable a user to defend against a SIA. In this scheme, the user (via their browser) inserts "fake" page visits into their browsing session. For example, the browser could make requests to different web pages in the background of a user's browsing. Though a user wouldn't see the content of the page, to any adversary it would appear as if the user were visiting that page.  This would have the effect of obfuscating a user's content fingerprint, as the fake page visits would misrepresent their profile. If an adversary cannot extract a meaningful content fingerprint for a user, then it will be difficult to correlate a user's sessions. Unfortunately, making random page visits is not sufficient, as an adversary might be able to normalize out random noise from a user's fingerprint. However, if a user's session's content fingerprint is made to look more like another user's session's content fingerprint, the adversary is less likely to successfully fingerprint either user. Thus, a user can defend themselves by adding page visits from other users' browsing sessions. 

We implement this defense as follows. For a single session $s_i$ belonging to a user $u$, we select several other sessions belonging to different users. For each session, we select a proportion of the pages and artificially add them to our original session $s_i$. In effect, this makes $s_i$ look more like these other sessions. We repeat this procedure for all sessions.

\begin{table}[h]
\centering
\begin{tabular}{lll}
Page Category & Frequency & Proportion \\
\hline \\
front page & 940469& 0.200 \\
news &452387 & 0.096 \\
tech &207479 &0.044 \\
local &386217 &0.082 \\
opinion &151409& 0.032 \\
on-air &414928& 0.088 \\
misc& 305615& 0.065 \\
weather &439398& 0.093 \\
health &196614& 0.041 \\
living &131760& 0.028 \\
business &96817& 0.020 \\sports &264899& 0.056 \\
summary& 216125& 0.045 \\
bulletin board service &395880& 0.084 \\
travel &56576& 0.012 \\
msn-news  &25249& 0.005 \\
msn-sports &16972& 0.003 \\
\end{tabular}
\caption{Distribution of page views over page categories}
\label{page_cat_dist}
\end{table}

\begin{table}[h]
\centering
    \begin{tabular}{ll}
    Top Tags      & Top Domains        \\\hline
    design        & www.youtube.com    \\
    tools         & www.guardian.co.uk \\
    education     & en.wikipedia.org   \\
    web           & mashable.com       \\
    web20         & www.nytimes.com    \\
    video         & www.slideshare.net \\
    blog          & github.com         \\
    webdesign     & lifehacker.com     \\
    viapackratius & d.hatena.ne.jp     \\
    inspiration   & www.bbc.co.uk      \\
    \end{tabular}
    \caption{Most popular domains and tags for Delicious data}
    \label{del_tags} 
\end{table}

\section{Results}
We now outline the results achieved by applying these techniques to the Delicious and MSNBC datasets. We begin by describing salient characteristics of both datasets. 

\subsection{Data Characteristics}
Most of the page visits in the MSNBC dataset are to "news",  "on-air" or "sports" pages (table \ref{page_cat_dist}). In the Delicious dataset, we see that most of the tags are technology related and that most of the commonly visited sites are news.

\subsection{Session Creation}
The MSNBC data consists of the page views (labeled by page category) from a user over the course of 24 hours. However, for a single user it collates all visits over the 24 hours, and doesn't distinguish between different sessions launched by that user. For a given page visit, we cannot determine when the user visited the page, or what other pages the user also visited in that session. Thus, we have to simulate separating the page visits into different sessions. We do this by randomly splitting each user's visited pages into two sets, and treat each set as a separate browsing session. This also allows our browsing sessions to conform to the consistency assumption described in section 4.1. 

Similarly, the Delicious dataset consists of various web pages tagged by users with bookmarks. Since this is an approximation of the browsing of each user (a user is unlikely to tag all the web pages they visit), it doesn't represent actual session based browsing activity. Like with the MSNBC dataset, we have to artificially create browsing sessions. Again, we do this by randomly splitting the pages tagged by a user into two sets, treating each as a distinct browsing session. 

\subsection{Attack Performance}

\subsubsection{Metric Methodology}
As discussed previously, different attackers may employ different definitions of "success", based on whether they prefer a strategy that maximizes precision or recall. In evaluating different attack strategies, we report the precision and recall  at the point at which the F1 score for the attack is maximized. This allows us to quantify how well the approach would work for the "average" adversary. In addition, we report the "reach" for each attack. This is the number of users successfully identified. 

When evaluating an attack strategy, we simulate the attack multiple times. In every trial, we sample a new set of users from our data. Since the success of a SIA attack is a function of the distinguishability of the users,  simulating multiple trials allows us to evaluate the range of attack performances. Since this is a privacy problem, we are primarily concerned with the worst case scenario. We therefore report the results of the worst case iteration. This is the trial where the adversary performs the best (the iteration with the highest F1 score). However, data regarding the average performance of each attack strategy across all iterations can be found in the appendix.   

\subsubsection{Baseline}
The baseline approach is equivalent to an adversary randomly guessing pairs of sessions to correspond to the same user.  We simulate this by assigning a random score between 0 and 1 for every pair of sessions. We designate all session pairs whose score exceeds a cutoff to belong to the same user. We report the precision and recall of this approach at the cutoff that maximizes the F1 score. 

\subsubsection{Attack Strategy Performances}
Table \ref{table:delicious_methods_worst}  and Table \ref{table:msnbc_methods_worst} contains the results of baseline, weighted closeness, and supervised learning attack strategies on the Delicious and MSNBC datasets respectively. For the weighted closeness strategy, we sampled 20 users with at least 40 page visits (at least 20 pages per session), and calculated the weighted closeness scores between all pairs of sessions. We then selected a cutoff that maximized the F1 score for the attack. We repeated this process 20 times for each attack, and report the results from the worst case results (when the adversary performs the best) in Table \ref{table:delicious_methods_worst}  and Table \ref{table:msnbc_methods_worst}. Table \ref{table:delicious_methods_average} and Table \ref{table:msnbc_methods_avg} in the Appendix contains the averaged performance over all 20 trials. On the Delicious data, we ran the weighted closeness strategy on page tag, domain, and page text fingerprint representations. On the MSNBC data, we ran the weighted closeness strategy on the page label fingerprint representation.  

For the supervised learning approach, we trained a neural network with a single hidden layer with 100 neurons on all session pairs from a sample of 20 users (where each user had 40 page visits). We then tested the network on all pairs of sessions from another sample of 20 users. We treated the predicted class probability for each pair of sessions as the similarity score between both sessions. We then selected a cutoff that maximized the F1 score for the attack. 

On the Delicious dataset, every approach except for the neural network (when applied to tags and domains) exceeded the baseline F1 score. Weighted closeness applied to page tags was the most successful strategy, correctly identifying sessions for 9 users (out of 20) with a  precision of 0.82 and a recall of 0.45. In general, we see that weighted closeness outperformed the neural network. 

On the MSNBC data, both the neural network and the weighted closeness approach exceeded the baseline F1 score. Like the Delicious data, the weighted closeness approach was the most successful, correctly identifying 14 users at a precision of 0.35 and a recall of 0.7. In comparing the results of the MSNBC data to the Delicious data, we can see that the weighted closeness was a more effective attack on the Delicious dataset. We can attribute this to the granularity of the Delicious page tags, which were far more descriptive than the MSNBC tags. The success of both approaches on the MSNBC data - in spite of the broadness of the MSNBC page labels - is indicative of the strength of both attacks.

\begin{table*}[!htbp]
\centering
\begin{tabular}{lllll}
Method &  F1 &  Precision &  Recall  &  Reach\\\hline 
Weighted closeness on Domain &  0.5 & 0.67 & 0.4 & 8\\ 
Weighted closeness on Tag  & 0.58 & 0.82 & 0.45 & 9\\ 
Weighted closeness on Text & 0.41 & 0.42 & 0.4 & 8\\ 
Neural Net on Tags &0.08&0.33&0.05	&1\\
Neural Net on Domain &	0.06&	0.05&	0.1	&2 \\
Neural Net on Words &0.17&0.15&0.2&4 \\
Baseline & 0.13&0.13&,0.15&,3\\
\end{tabular}
\caption{Performance of various attacks for the Delicious dataset} 
\label{table:delicious_methods_worst}
\end{table*}

We also investigated the effects of the user sample size on the performances of these attacks. Instead of running the attack on a fixed sample of 20 users, we experimented on different numbers of users.  Table \ref{table:msnbc_num_users_worst},\ref{table:msnbc_num_users_avg},\ref{table:delicious_num_users_avg}, and \ref{table:delicious_num_users_worst}  (found in the Appendix A) contain the results for different user sample sizes on the MSNBC and Delicious data respectively. For each sample size, we ran 20 iterations of a weighted closeness attack (using page tags for Delicious data and the page labels for MSNBC data) and reported the results from the worst case trial and the average across all trials. As we can see, the performance of the attacks steadily decrease as the number of users increases. This result is expected as identifying sessions becomes harder when the number of users increases and users become harder to distinguish. However, we can see that even with a 45 user sample, we are able to identify 19 users with an overall precision of 0.5 and a recall of 0.47.

We also investigated the effect of session length. Table  \ref{table:msnbc_pps_worst},\ref{table:msnbc_pps_avg}, \ref{table:delicious_pps_worst} , and \ref{table:delicious_pps_avg} (both found in the Appendix) contain the results for different minimum session lengths on the MSNBC and Delicious data respectively. For each sample size, we ran 20 iterations of a weighted closeness attack (using page tags for Delicious data and the page labels for MSNBC data) and reported the results from the worst case trial and the average across all trials. As we can see, the performance of the attacks steadily increases as the minimum session length increases. This result is expected as the content fingerprint for sessions become more representative of the user browsing profile as the session grows in length.

\begin{table*}[!htbp]
\centering
\begin{tabular}{lllll}
Method &  F1 & Precision &  Recall  &  Reach\\\hline
Weighted closeness  & 0.47 & 0.35 & 0.7 & 14\\ 
Neural  &  0.25 & 0.33 & 0.2 & 2\\ 
Baseline &  0.13 & 0.2 & 0.1 & 2\\ 
\end{tabular}
\caption{Performance of different type of attacks on the MSNBC Dataset}
\label{table:msnbc_methods_worst}
\end{table*}

\subsection{Defense Performance}
We also evaluated the strength of our defensive mechanisms. We validated these countermeasures by running them on a set of user sessions and launching a weighted closeness attack on the sessions.  We sampled 20 users with at least 40 page visits (at least 20 pages per session), and implemented the obfuscation mechanisms. We then launched a weighted closeness attack (using the page tags as features for the delicious data). We repeated this process 20 times. In Table \ref{table:delicious_defense_worst} and Table \ref{table:msnbc_defense_worst} we can see the results of the defenses. "Session sample size" is the number of other sessions we randomly sampled and "Page sample size" is the number of pages we sampled from each of these sessions to add to the user's browsing.
As we can see, the defense worked best for both Delicious and MSNBC when we sampled 10 pages from a single random session. For the average results of these trials, see Table \ref{table:delicious_defense_avg} and Table \ref{table:msnbc_methods_avg} in the Appendix.
 
\begin{table*}[h]
\centering
\begin{tabular}{llllll}
Session sample size  & Page sample size & F1 &  Precision &  Recall  &  Reach\\ \hline \\
1 & 5 & 0.49 & 0.48 & 0.5 & 10\\ 
1 & 10 &  0.32 & 0.45 & 0.25 & 5\\ 
2 & 5 & 0.5&0.56&	0.45&	9\\
\end{tabular}
\caption{Performance of various defensive measures on the Delicious dataset (worst case trial)} 
\label{table:delicious_defense_worst}
\end{table*}

 \begin{table*}[h]
\centering
\begin{tabular}{llllll}
Session Sample Size  & Page Sample Size & F1 & Precision & Recall  & Reach\\ \hline \\
2 & 5  & 0.41 & 0.32 & 0.55 & 11 \\
1 & 5   & 0.48&0.77&0.35&7.0 \\ 
1 & 10  & 0.32 & 0.45 & 0.25 & 5 \\ 
\end{tabular}
\caption{Performance of various defenses on MSNBC dataset (worst case trial)}
\label{table:msnbc_defense_worst}
\end{table*}

\section{Discussion}

Our work demonstrates the feasibility of a semantic identification attack and the threat it poses to privacy. The attack strategies we present were demonstrated to be effective, capable of identifying, for a significant fraction of users,  when two sessions belong to the same user. We show that even with coarse signals - such as those present in the MSNBC data - we are capable of correctly associating a significant fraction of users' sessions. 

On the Delicious dataset, nearly all methods exceeded the baseline approach. Computing weighted closeness distance scores  on web page tags was the most effective attack, performing better on average (and in the worst case) than all other methods. The success of all three types of features offer insight into user behavior. The success of the domains feature implies that users tend to return to the same domains, but differ from each other in the domains they visit. The success of the page tags supports the intuition behind the SIA - that user browsing is  topically consistent and motivated by a fixed set of interests. Finally, the success of the page text feature demonstrates that users are relatively consistent in the content they consume.  

In general, the weighted closeness distance proved to be more effective at identifying users than the neural net approach. Amongst the features, the weighted closeness  approach was most successful when using page tags.   The relative success of page tags (over page content and page domains) highlights the importance of granularity for semantic signals. If the signals are too coarse, then the generated fingerprint will not be specific enough and sessions from different users will have too similar fingerprints. In contrast, if the semantic signal is too specific, the fingerprints for each session will be too different. In this case, we will likely predict sessions from the same user as originating from different users.  In the case of the Delicious dataset, we see that tag features present an optimal level of granularity between the page domains and page text. However, we could also interpret this to mean that our text model was unable to successfully capture the core topics of each page. This is something we hope to investigate in future work.

On the MSNBC dataset, the weighted closeness distance metric also performs the best. However, it's  performance is significantly worse than the  performance of the weighted closeness distance method on the Delicious dataset. This is likely because the Delicious tags are far richer and convey more semantic information than the MSNBC page labels.  

Intuitively, it makes sense that the performance of an attack would decrease as the sample of users increases in size. We expect a majority of users to be relatively indistinguishable. As our sample increases in size, the number of indistinguishable users will grow disproportionately, resulting in the attack misclassifying a greater number of session pairs. In both the MSNBC and Delicious datasets, we see that as the number of users in the sample increases, the performance of the algorithm reduces significantly. As we discussed previously, this result is expected. The problem of distinguishing users is highly dependent on the number of users. As the sample size increases, users become harder to distinguish.  

The defense mechanism we formulated represents an improvement in privacy for users. On both the Delicious and MSNBC datasets, the defensive measures were successful in lowering the F1 score for the attack.   

The range of potential semantic signals extends far beyond those referenced (or used) in this paper. For example, they could use third party cookies to gather the pages visited by a user across multiple sites \cite{mayer2012third}. Alternatively, a first party website could log every action by a user on its website and use these to construct site specific content fingerprints. In either case, the adversary would have access to a set of semantic signals significantly richer than those present in our data. In the best case for users, the adversary would merely replicated our performance. In the worst case, they might fare significantly better and identify more users.

In summary, the results of our experiments demonstrate that an adversary employing our attacks could successfully extract content fingerprints from a subset of sessions and use those fingerprints to identify when two sessions were launched by the same user. We presented a framework for evaluating the success of these attacks in the context of an adversary's goals and demonstrate how different attacks could be advantageous to different adversaries. 

It is important to recognize that semantic identification attacks are applicable to a vastly greater range of contexts than those presented in this paper. The New York Times for example, may run a semantic identification attack on each of its registered accounts to see if multiple individuals are using the same account. A shopping website may launch an attack to identify users based on their interests or habits.  

Going beyond web browsing, they can be generalized to any kind of behavior pattern. They allow an adversary to launch an identification attack on an individual's habits and routines, from driving routes to purchasing habits. As these are significantly harder to disguise than metadata, semantic identification attacks can be incredibly powerful.
\section{Acknowledgments}
We'd like to thank Ramakrishnan Srikant, Dan Boneh, Ramanathan Guha, and Mehran Sahami,  Lea Kissner, Scott Ellis, and Jonathan Mayer for their advice, and guidance on this project. 
\bibliographystyle{abbrv}
\bibliography{sources}

\begin{thebibliography}{10}

\bibitem{pano}
\url{https://panopticlick.eff.org/}.

\bibitem{amiunique}
\url{https://amiunique.org/}.

\bibitem{msnbc_data}
\url{https://archive.ics.uci.edu/ml/datasets/MSNBC.com+Anonymous+Web+Data}.

\bibitem{delicious}
\url{https://grouplens.org/datasets/hetrec-2011/}.

\bibitem{scikit}
\url{http://scikit-learn.org/stable/modules/generated/sklearn.neural_network.MLPClassifier.html#sklearn.neural_network.MLPClassifier}.

\bibitem{rsa}
S.~S. Africa.
\newblock General household survey.

\bibitem{agrawal2000privacy}
R.~Agrawal and R.~Srikant.
\newblock Privacy-preserving data mining.
\newblock {\em ACM Sigmod Record}, 29(2):439--450, 2000.

\bibitem{Dwork2008}
C.~Dwork.
\newblock {\em Differential Privacy: A Survey of Results}, pages 1--19.
\newblock Springer Berlin Heidelberg, Berlin, Heidelberg, 2008.

\bibitem{Eckersley2010}
P.~Eckersley.
\newblock {\em How Unique Is Your Web Browser?}, pages 1--18.
\newblock Springer Berlin Heidelberg, Berlin, Heidelberg, 2010.

\bibitem{goel2012does}
S.~Goel, J.~M. Hofman, and M.~I. Sirer.
\newblock Who does what on the web: A large-scale study of browsing behavior.
\newblock 2012.

\bibitem{5708146}
M.~Gotz, A.~Machanavajjhala, G.~Wang, X.~Xiao, and J.~Gehrke.
\newblock Publishing search logs- a comparative study of privacy guarantees.
\newblock {\em IEEE Transactions on Knowledge and Data Engineering},
  24(3):520--532, March 2012.

\bibitem{aol_scandal}
S.~Hansell.
\newblock Aol removes search data on vast group of web users.
\newblock August 2008.

\bibitem{Jones:2007:IKY:1321440.1321573}
R.~Jones, R.~Kumar, B.~Pang, and A.~Tomkins.
\newblock "i know what you did last summer": Query logs and user privacy.
\newblock In {\em Proceedings of the Sixteenth ACM Conference on Conference on
  Information and Knowledge Management}, CIKM '07, pages 909--914, New York,
  NY, USA, 2007. ACM.

\bibitem{Kumar:2007:AQL:1242572.1242657}
R.~Kumar, J.~Novak, B.~Pang, and A.~Tomkins.
\newblock On anonymizing query logs via token-based hashing.
\newblock In {\em Proceedings of the 16th International Conference on World
  Wide Web}, WWW '07, pages 629--638, New York, NY, USA, 2007. ACM.

\bibitem{mayer2012third}
J.~R. Mayer and J.~C. Mitchell.
\newblock Third-party web tracking: Policy and technology.
\newblock In {\em 2012 IEEE Symposium on Security and Privacy}, pages 413--427.
  IEEE, 2012.

\bibitem{mowery2011fingerprinting}
K.~Mowery, D.~Bogenreif, S.~Yilek, and H.~Shacham.
\newblock Fingerprinting information in javascript implementations.
\newblock {\em Proceedings of W2SP}, 2:180--193, 2011.

\bibitem{MS12}
K.~Mowery and H.~Shacham.
\newblock Pixel perfect: Fingerprinting canvas in {HTML5}.
\newblock In M.~Fredrikson, editor, {\em Proceedings of W2SP 2012}. IEEE
  Computer Society, May 2012.

\bibitem{Narayanan:2008:RDL:1397759.1398064}
A.~Narayanan and V.~Shmatikov.
\newblock Robust de-anonymization of large sparse datasets.
\newblock In {\em Proceedings of the 2008 IEEE Symposium on Security and
  Privacy}, SP '08, pages 111--125, Washington, DC, USA, 2008. IEEE Computer
  Society.

\bibitem{narayanan2010myths}
A.~Narayanan and V.~Shmatikov.
\newblock Myths and fallacies of personally identifiable information.
\newblock {\em Communications of the ACM}, 53(6):24--26, 2010.

\bibitem{Novak:2004:AW:988672.988678}
J.~Novak, P.~Raghavan, and A.~Tomkins.
\newblock Anti-aliasing on the web.
\newblock In {\em Proceedings of the 13th International Conference on World
  Wide Web}, WWW '04, pages 30--39, New York, NY, USA, 2004. ACM.

\bibitem{china_ic}
J.~L. Qiu.
\newblock Cybercafés in china: Community access beyond gaming and tight
  government control.
\newblock 2013.

\bibitem{rivest1998chaffing}
R.~L. Rivest et~al.
\newblock Chaffing and winnowing: Confidentiality without encryption.
\newblock {\em CryptoBytes (RSA laboratories)}, 4(1):12--17, 1998.

\bibitem{Yen2009}
T.-F. Yen, X.~Huang, F.~Monrose, and M.~K. Reiter.
\newblock {\em Browser Fingerprinting from Coarse Traffic Summaries: Techniques
  and Implications}, pages 157--175.
\newblock Springer Berlin Heidelberg, Berlin, Heidelberg, 2009.

\bibitem{host-fingerprinting-and-tracking-on-the-webprivacy-and-security-implications}
T.-F. Yen, Y.~Xie, F.~Yu, R.~P. Yu, and M.~Abadi.
\newblock Host fingerprinting and tracking on the web:privacy and security
  implications.
\newblock In {\em The 19th Annual Network and Distributed System Security
  Symposium (NDSS) 2012}. Internet Society, February 2012.

\end{thebibliography}

\onecolumn
\appendix

\begin{table*}[h]
\caption{Results of various attacks for the Delicious dataset (averaged across all trials)} 
\begin{tabularx}{\textwidth}{XXXXX}
Method & F1 & Precision & Recall &Reach\\ 
\hline
Weighted Closeness (domains) & 0.35 & 0.46 & 0.32 & 6.5 \\ 
Weighted Closeness (tags) & 0.48 & 0.56 & 0.48 & 9.6 \\ 
Weighted Closeness (text) & 0.25 & 0.3 & 0.27 & 5.3 \\ 
Neural Net (tags) & 0.052&	0.041&0.95&19.05\\
Neural Net (domains) & 0.054&	0.03&	0.75&	15\\
Neural Net (words) &0.16&0.18&0.19&3.8\\
Baseline&0.073&0.054&0.47&9.45\\
\end{tabularx}
\label{table:delicious_methods_average}
\end{table*}

\begin{table*}[h]
\caption{Performance of different type of attacks on the MSNBC Dataset (averaged across all trials)} 
\centering
\begin{tabularx}{\textwidth}{XXXXX}
Method & Average F1 & Average Precision & Average Recall & Average Reach \\ 
\hline
Weighted closeness (labels) & 0.29 & 0.60 & 0.31 & 6.35 \\ 
Neural (labels) & 0.11 & 0.13 & 0.74 & 7.4 \\ 
Baseline & 0.08 & 0.07 & 0.25 & 4.9 \\ 
\end{tabularx}
\label{table:msnbc_methods_avg}
\end{table*}

\begin{table*}[h]
\caption{Results of various defensive measures on the Delicious dataset (averaged across all trials)} 
\centering
\begin{tabularx}{\textwidth}{XXXXXX}
Session sample size  & Page sample size & F1 & Precision & Recall & Reach \\ \hline \\
1 & 5 & 0.27 & 0.30 & 0.34 & 6.75 \\ 
1 & 10 & 0.17 & 0.19 & 0.24 & 4.8 \\ 
2 & 5 & 0.21&	0.23&0.27&5.4\\
\end{tabularx}
\label{table:delicious_defense_avg}
\end{table*}

 \begin{table*}[h]
 \caption{Results of various defenses on MSNBC dataset (averaged across all trials)}
\centering
\begin{tabularx}{\textwidth}{XXXXXX}
Session sample size  & Pages Sample Size & F1 & Precision & Recall & Reach \\ \hline\\
2 & 5 & 0.25 & 0.22 & 0.39 & 7.75 \\
1 & 5 & 0.30&,0.33&0.42&8.55 \\ 
1 & 10 & 0.21 & 0.22 & 0.28 & 5.65 \\ 
\end{tabularx}
\label{table:msnbc_defense_avg}
\end{table*}

\begin{table*}[h]
\caption{Results of different minimum session lengths on MSNBC dataset (worst case trial)} 
\begin{tabularx}{\textwidth}{XXXXX}
Minimum Session Length &  F1 &  Precision &  Recall  &  Reach\\ 
\hline
5 & 0.52 & 1 & 0.35 & 7\\ 
10 & 0.48 & 0.78 & 0.35 & 7\\ 
15 &  0.52 & 0.5 & 0.55 & 11\\ 
20 & 0.47 & 0.36 & 0.7 & 14\\ 
25 & 0.59 & 0.65 & 0.55 & 11\\ 
30 &  0.58 & 0.5 & 0.7 & 14\\ 
35 & 0.55 & 0.52 & 0.6 & 12\\ 
40 &  0.73 & 0.92 & 0.6 & 12\\ 
\end{tabularx}
\label{table:msnbc_pps_worst}
\end{table*}

\begin{table*}[h]
\caption{Results of different minimum session lengths on MSNBC dataset (averaged across all trials)} 
\begin{tabularx}{\textwidth}{XXXXX}
Minimum Session Length &  F1 &  Precision &  Recall &  Reach \\
\hline
5 & 0.3 & 0.53 & 0.3 & 5.9 \\
10 & 0.28 & 0.56 & 0.22 & 4.4 \\
15 & 0.33 & 0.56 & 0.34 & 6.85 \\
20 & 0.29 & 0.60 & 0.32 & 6.35 \\
25 & 0.31 & 0.53 & 0.35 & 7.05 \\
30 & 0.34 & 0.46 & 0.44 & 8.85 \\
35 & 0.32 & 0.46 & 0.48 & 9.65 \\
40 & 0.35 & 0.59 & 0.41 & 8.15 \\
\end{tabularx}
\label{table:msnbc_pps_avg}
\end{table*}

 \begin{table*}[h]
\centering
\caption{Effect of minimum session lengths on Delicious dataset (worst case trials)} 
\begin{tabularx}{\textwidth}{XXXXX}
Minimum Session Length &   F1 &  Precision &  Recall  &  Reach\\ 
\hline
5 &  0.67 & 0.85 & 0.55 & 11\\ 
10 &  0.67 & 0.85 & 0.55 & 11\\ 
15 & 0.71 & 0.86 & 0.6 & 12\\ 
20 &  0.76 & 0.82 & 0.7 & 14\\ 
25 &  0.72 & 0.6 & 0.9 & 18\\ 
\end{tabularx}
\label{table:delicious_pps_worst}
\end{table*}

 \begin{table*}[h]
\centering
\caption{Results of different minimum session lengths on Delicious dataset (averaged across all trials)} 
\begin{tabularx}{\textwidth}{XXXXX}
Minimum Session Length & F1 &  Precision &  Recall &  Reach \\
5 & 0.48 & 0.58 & 0.44 & 8.75 \\
10 & 0.50 & 0.54 & 0.515 & 10.3 \\
15 & 0.51 & 0.55 & 0.50 & 9.9 \\
20 & 0.51 & 0.57 & 0.53 & 10.55 \\
25 & 0.49 & 0.59 & 0.51 & 10.25\\
\end{tabularx}
\label{table:delicious_pps_avg}
\end{table*}

 \begin{table*}[h]
 \caption{Results of different user sample sizes on Delicious dataset (worst case trials)} 
\begin{tabularx}{\textwidth}{XXXXX}
Number of Users &  F1 & Precision & Recall  & Reach\\ 
\hline
5 &  1 & 1 & 1 & 5\\ 
10 &  0.76 & 0.72 & 0.8 & 8\\ 
15 &  0.65 & 0.52 & 0.86 & 13\\ 
20 & 0.70 & 0.85 & 0.6 & 12\\ 
25 &  0.61 & 0.62 & 0.6 & 15\\ 
30 &  0.64 & 0.73 & 0.56 & 17\\ 
35 &  0.47 & 0.47 & 0.48 & 17\\ 
40 &  0.48& 0.5 & 0.47 & 19\\ 
\end{tabularx}
\label{table:delicious_num_users_worst}
\end{table*}
 
   \begin{table*}[h]
 \caption{Results of different user sample sizes on Delicious dataset (averaged across all trials)} 
\begin{tabularx}{\textwidth}{XXXXX}
Number of Users &  F1 &  Precision & Recall & Reach \\ 
\hline
5 & 0.75 & 0.79 & 0.78 & 3.9 \\ 
10 & 0.57 & 0.65 & 0.56 & 5.6 \\ 
15 & 0.52 & 0.63 & 0.5 & 7.5 \\ 
20 & 0.46 & 0.52 & 0.47 & 9.55 \\ 
25 & 0.43 & 0.49 & 0.44 & 11.2 \\
30 & 0.42 & 0.46 & 0.42 & 12.75 \\
35 & 0.35 & 0.39 & 0.36 & 12.75 \\
40 & 0.35 & 0.38 & 0.36 & 14.45 \\
\end{tabularx}
\label{table:delicious_num_users_avg}
\end{table*}

 \begin{table*}[h]
\caption{Results of different user sample sizes on MSNBC dataset (worst case trials)} 
\begin{tabularx}{\textwidth}{XXXXX}
User Sample Size  &  F1 &  Precision &  Recall  &  Reach\\ 
\hline
5 &  1 & 1 & 1 & 5\\ 
10  & 0.77 & 0.625 & 1 & 10\\ 
15 &  0.61 & 0.52 & 0.73 & 11\\ 
20 &  0.58 & 0.61 & 0.55 & 11\\ 
25 &  0.45 & 0.43 & 0.48 & 12\\ 
30 &  0.42 & 0.69 & 0.3 & 9\\ 
35 &  0.52 & 0.48 & 0.57 & 20\\ 
40 &  0.37 & 0.35 & 0.4 & 16\\ 
\end{tabularx}
\label{table:msnbc_num_users_worst}
\end{table*}

 \begin{table*}[h]
\caption{Results of different user sample sizes on MSNBC dataset (averaged across all trials)} 
\begin{tabularx}{\textwidth}{XXXXX}
User Sample Size &  F1 &  Precision &  Recall &  Reach\\
\hline
5 & 0.60 & 0.70 & 0.69\\
10 & 0.46 & 0.54 & 0.60 & 5.95 \\
15 & 0.34 & 0.57 & 0.38 & 5.75 \\
20 & 0.35 & 0.65 & 0.35 & 7.05 \\
25 & 0.25 & 0.56 & 0.25 & 6.3 \\
30 & 0.21 & 0.49 & 0.32 & 9.45 \\
35 & 0.25 & 0.52 & 0.21 & 7.4 \\
40 & 0.19 & 0.54 & 0.25 & 9.85 \\
\end{tabularx}
\label{table:msnbc_num_users_avg}
\end{table*}
\end{document}